\documentclass{webofc}

\usepackage[varg]{txfonts}   
\usepackage{hyperref}
\usepackage{url}
\hypersetup{colorlinks=true,citecolor=blue,urlcolor=blue,linkcolor=blue}
\begin{document}
\title{LHCb Stripping Project: Continuing to Fully and Efficiently Utilize Legacy Data}
%
%

\author{\firstname{Nathan} \lastname{Grieser}\inst{1}\fnsep\thanks{\email{ngrieser@cern.ch}} \and
        \firstname{Federico Leo} \lastname{Redi}\inst{2}\fnsep \and
        \firstname{Eduardo} \lastname{Rodrigues}\inst{3}
        \and
        \firstname{Niladri} \lastname{Sahoo}\inst{4}
        \and
        \firstname{Shuqi} \lastname{Sheng}\inst{5,}\inst{6}
        \and
        \firstname{Nicole} \lastname{Skidmore}\inst{7}
        \and
        \firstname{Mark} \lastname{Smith}\inst{8}
        \and
        \firstname{Aravind} \lastname{Venkateswaran}\inst{9}
}

\institute{University of Cincinnati (US)
\and
           Università degli Studi e INFN Milano (IT)
\and
           University of Liverpool (GB)
\and
            University of Birmingham (GB)
\and
            Institute of High Energy Physics (CN)
\and
            University of Chinese Academy of Sciences (CN)
\and
            University of Warwick (GB)
\and
            Imperial College London (GB)
\and
            EPFL - Ecole Polytechnique Federale Lausanne (CH)
          }

\abstract{The LHCb collaboration continues to heavily utilize the Run 1 and Run 2 legacy datasets well into Run 3. As the operational focus shifts from the legacy data to the live Run 3 samples, it is vital that a sustainable and efficient system is in place to allow analysts to continue to profit from the legacy datasets. The LHCb Stripping project is the user-facing offline data-processing stage that allows analysts to select their physics candidates of interest simply using a Python-configurable architecture. After physics selections have been made and validated, the full legacy datasets are then reprocessed in small time windows known as Stripping campaigns.

Stripping campaigns at LHCb are characterized by a short development window with a large portion of collaborators, often junior researchers, directly developing a wide variety of physics selections; the most recent campaign dealt with over 900 physics selections. Modern organizational tools, such as GitLab Milestones, are used to track all of the developments and ensure the tight schedule is adhered to by all developers across the physics working groups. Additionally, continuous integration is implemented within GitLab to run functional tests of the physics selections, monitoring rates and timing of the different algorithms to ensure operational conformity. Outside of these large campaigns the project is also subject to nightly builds, ensuring the maintainability of the software when parallel developments are happening elsewhere.
}
\maketitle
\clearpage
\section{Introduction}
\label{intro}
As high-energy physics experiments evolve, so too do their dataflows and data models. The LHCb collaboration continues to rely heavily on its Run 1 and Run 2 legacy datasets even as the focus shifts toward incoming data from Run 3. To ensure a continuously vibrant physics program, it is crucial to maintain a sustainable and efficient system that allows analysts to seamlessly access and utilize legacy data. Significant challenges arise in such an ecosystem: the software infrastructure must remain adaptable to evolving needs, be efficient in handling large-scale reprocessing campaigns, and accessible to a broad user base, including new collaborators.

A key component of this ecosystem is the LHCb Stripping project, which serves as the user-facing offline data-processing stage. This system enables analysts to select their physics candidates of interest using a Python-configurable architecture. Once selections are finalized and validated, the full legacy datasets are reprocessed in structured phases known as Stripping campaigns. These campaigns operate within tight development windows and involve a large, diverse group of contributors—many of whom are junior researchers—who actively develop hundreds of physics selections. To manage this complexity, modern organizational tools such as GitLab Milestones are employed to track developments, ensuring timely completion across physics working groups. Additionally, continuous integration (CI) is used to run functional tests, monitoring algorithm performance and maintaining operational reliability. Beyond these structured campaigns, nightly builds further ensure software stability amid parallel developments.

Given the dynamic nature of both data processing and collaborative development, a sustainable approach must prioritize adaptability, ease of use, and rigorous software validation. A simple, well-structured software framework enables efficient workflows that can quickly respond to operational changes, while regular testing helps track the impact of sporadic modifications. Furthermore, effective knowledge transfer is essential for long-term maintainability, allowing new collaborators to adopt and refine procedures as needed. The Stripping project fulfills the  goal to establish an easy-to-use, sustainable legacy workflow that ensures continued access to valuable physics datasets, while also supporting the evolving needs of the LHCb physics program.

\section{LHCb Legacy Dataflow and Software Structure}
\label{sec:Stack}

\begin{figure}[h]
\centering
\includegraphics[width=.99\linewidth,clip]{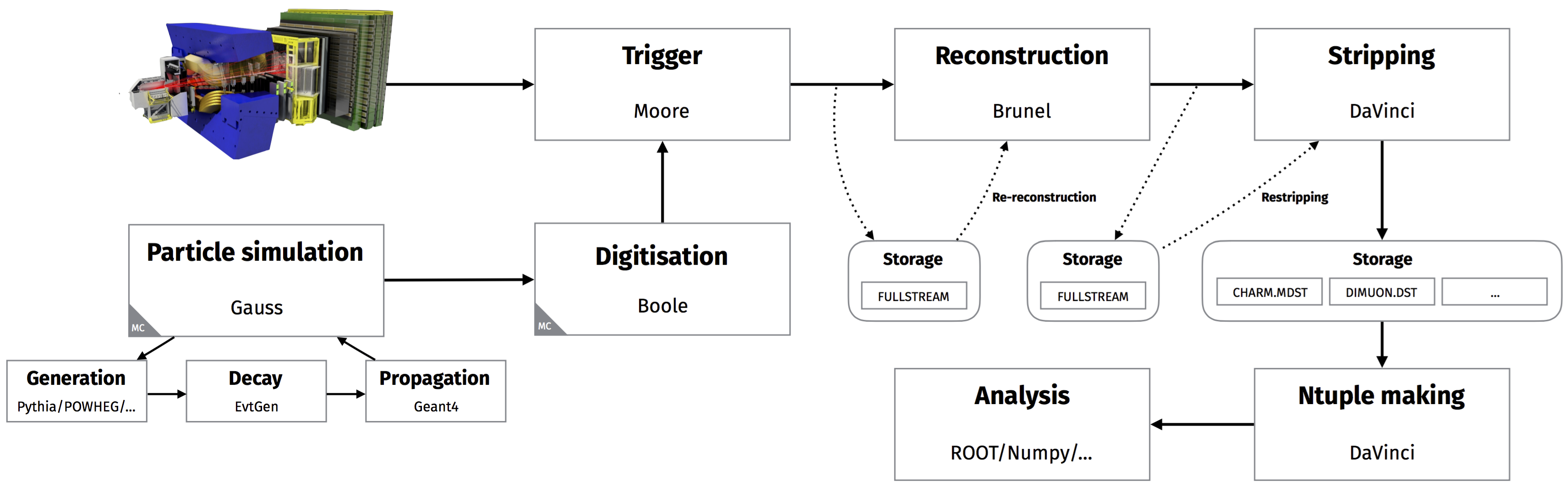}
\caption{LHCb Dataflow for LHC Runs 1 and 2.  Figure taken from Ref. \cite{CERN-LHCC-2018-007}.}
\label{fig:Dflow}       
\end{figure}

The LHCb experiment conducts a broad range of physics measurements, with a primary focus on studying charm and beauty decays. Since charm and beauty hadrons are abundantly produced in proton-proton collisions at the LHC, the data processing model faces two key challenges related to output bandwidth. First, events must be selected with high purity to ensure relevant data is retained. Second, the amount of event information available to analysts needs to be reduced for efficient processing. These challenges are addressed within the LHCb data processing model, which spans from event triggering to the final production of data files for analysis.  A diagram representing the dataflow for the LHC runs 1 and 2 is shown in Figure~\ref{fig:Dflow}.

\begin{figure}[h]
\centering
\includegraphics[width=5cm,clip]{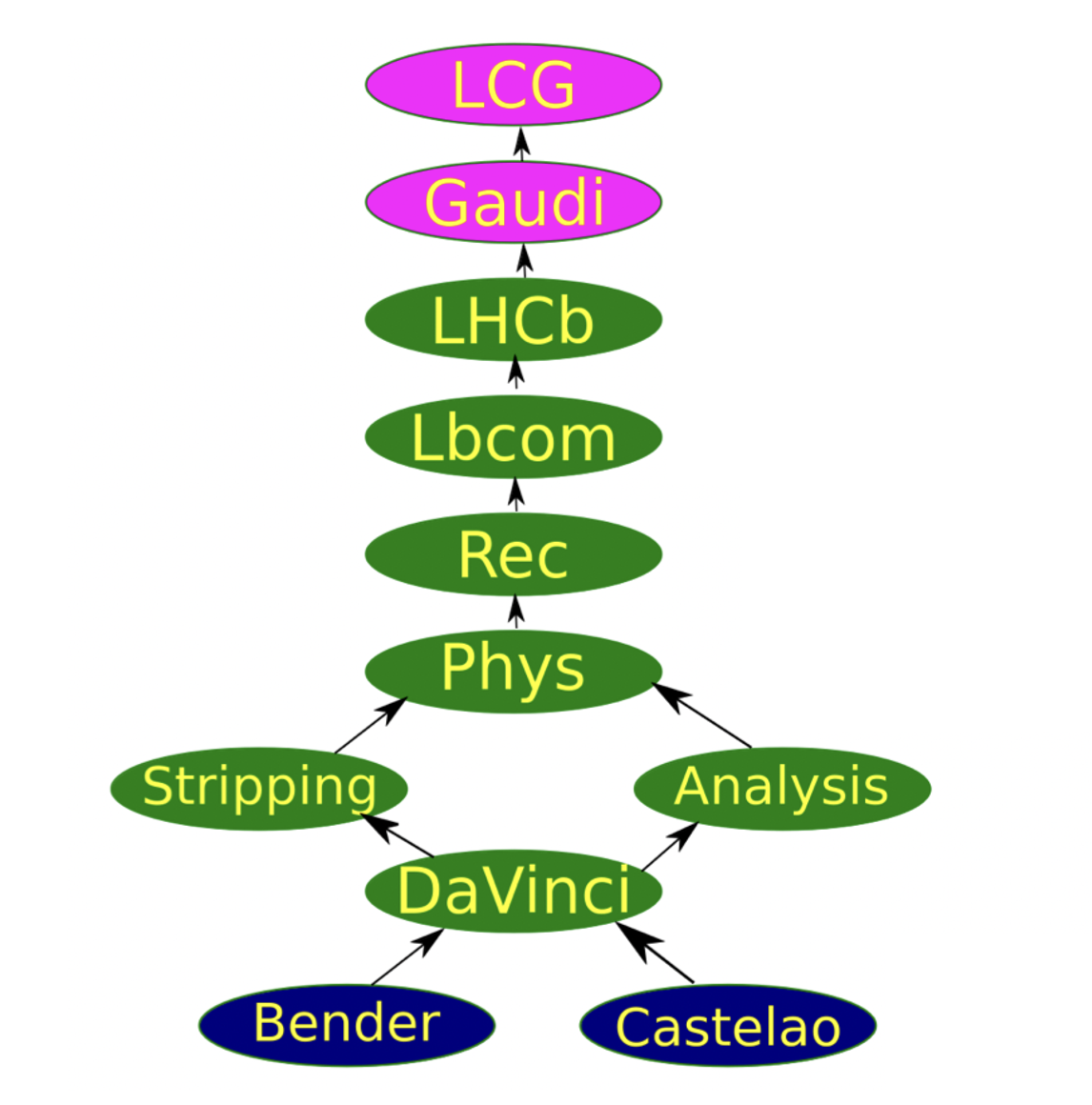}
\caption{THE LHCb software dependency hierarchy.  Green projects are managed by the LHCb collaboration, magenta projects are provided by the HEP community, and blue projects are LHCb managed projects who have been deprecated to ease sustainability.}
\label{fig:Stack}       
\end{figure}

The LHCb experiment maintains a dedicated legacy software stack for analyzing Run 1 and Run 2 data, visualized in Figure~\ref{fig:Stack}. This stack is continuously updated from LHCb~\cite{LHCb} project upwards, incorporating the latest LCG~\cite{LCG} and Gaudi~\cite{BARRAND200145} builds while discontinuing maintenance for obsolete projects\footnote{Projects refer to independently managed and releasable blocks of software.}. Stability and performance are ensured through collaboration with the core computing team. Additionally, new tools can be introduced to process legacy data after production, though releases are made only when necessary. The processing framework follows a structured hierarchy, with key components such as Stripping, Analysis, and DaVinci~\cite{DV} projects enabling efficient data selection and refinement, while ad hoc tools like Bender and Castelao provide further specialized capabilities.

The LHCb Stripping project is responsible for the first layer of offline data processing.  The Stripping project streamlines reconstructed collision data into a manageable subset for detailed physics analysis. Processed datasets are merged into 5 GB files, stored on disk, and replicated on tape, with data placement determined by available storage. As the initial user-facing offline data-processing stage, it allows analysts to select physics candidates through a Python-configurable architecture. Once selections are validated, legacy datasets are (re-)processed in ``(Re-)Stripping campaigns.”

During these campaigns, data collected by the LHCb detector is skimmed and slimmed to retain only the most relevant information for physics analysis. Selection criteria and algorithms identify key events and particles to be stored. To manage this complex process within tight timelines—often dictated by shutdown schedules—GitLab Milestones help track progress across Physics Working Groups (PWGs). Beyond large campaigns, nightly builds ensure software maintainability amid ongoing parallel developments.

\section{Performing a Stripping Campaign}

As the community of HEP looks towards the exciting future of the high luminosity LHC, older datasets will often take a back seat in priority of operational resources.  Therefore, it becomes critical that the developmental procedure is well defined, clear timelines are shared with management teams, and person-power is properly allocated to different tasks.

\subsection{Organization of Participants}

The primary responsibility of Stripping campaigns is placed on the Stripping Coordinators.  The Stripping coordination team is responsible for facilitating communication between the Operations Planning Group and the Physics Planning Group to ensure both operational and analysis constraints are met. Their duties include preparing necessary repositories and scripts, setting up validation procedures, reviewing merge requests (MRs), creating nightly tests, and managing production requests and documentation for Stripping campaigns.

\begin{figure}[h]
\centering
\includegraphics[width=.99\linewidth]{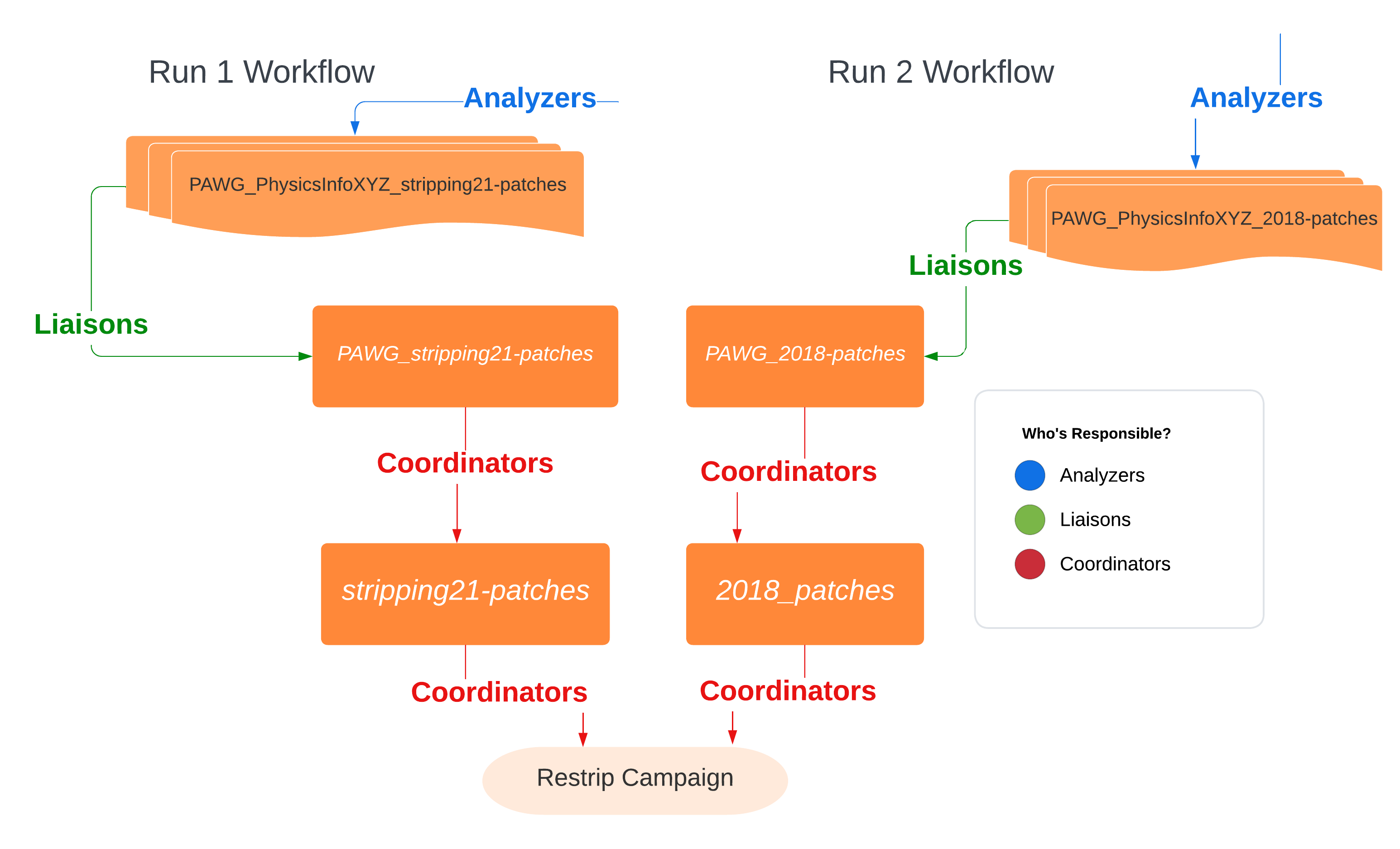}
\caption{A flow chart illustrating the development workflow for a Stripping campaign.  Each stage is identified with different responsibilities to ensure proper validation and performance each step of the developmental process.}
\label{fig:WFlow}       
\end{figure}

The second type of oversight participant in a campaign are PWG liaisons.  Each PWG is assigned a liaison responsible for coordinating Stripping-related tasks during campaign preparations. Their key duties include communicating essential information about deadlines, workflows, documentation, and data availability to their PWG, either directly or through conveners. They also assist members in writing new Stripping lines and validating updated ones through testing. Additionally, liaisons oversee MRs from analysts, ensuring proper integration into the development branch while maintaining successful CI test results.  A flow chart showing the breakdown of responsibilities during a development stage of a Stripping campaign is provided in Figure~\ref{fig:WFlow}.

\subsection{Determining Campaign Needs and Viability}

Large HEP experiments often have significant operational constraints which would not allow for a re-processing of full offline datasets at-will.  Gauging the physics needs of the collaboration is a responsibility carried out by the Stripping Coordinators in conjunction with the Physics Coordination team of LHCb.  The need for a large reprocessing campaign is primarily driven by analysts. Before initiating such a campaign, a survey is conducted to assess demand and specific interests in reprocessed data. Most responses typically focus on improving selections for ongoing analyses rather than addressing bug fixes. The collected feedback is then reviewed by the Physics Coordination team, which decides on the scope and focus of the campaign. 

\begin{figure}[h]
\centering
\includegraphics[width=.99\linewidth]{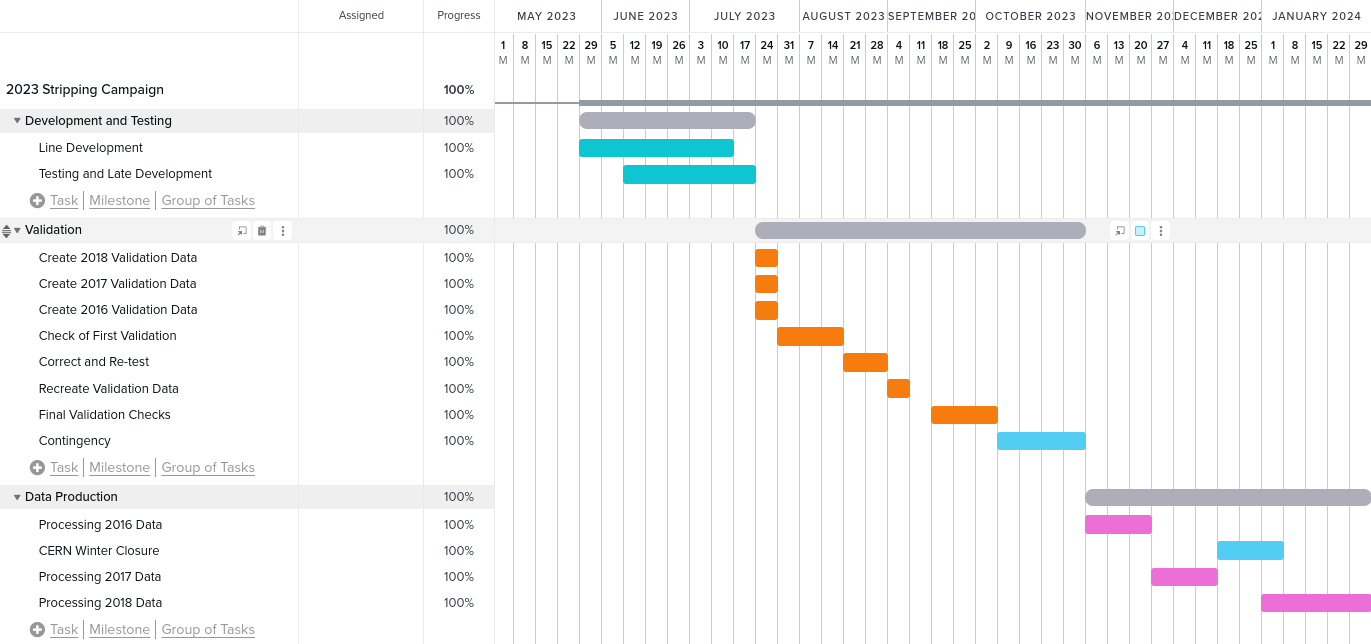}
\caption{A gantt chart illustrating the full proposed schedule for a Stripping campaign.  Clear timeline breakdowns are critical for cohesion between analysts, operational supporters, and management teams.}
\label{fig:Sched}       
\end{figure}

Once agreement is reached that a campaign would significantly benefit the experiment's physics program, interface between the Operations Planning Group is carried out by the Stripping Coordination team.  Discussions on timelines, resource availability, and data management schemes all play a role in the final decision of whether a Stripping campaign can be viable.  In the case of an affirmative decision, a detailed timeline is provided to analysts, distributed computing teams, and management to facilitate an efficient process.  An example of gantt chart highlighting the timeline of the most recent Stripping campaign is provided in Figure~\ref{fig:Sched}.

\subsection{Performing a Stripping Campaign}

Unfortunately, it is near impossible to continuously update a legacy software stack to be identical to the procedures of the ``live" stack of a given HEP experiment.  Therefore, before a campaign can begin, a training event must be held to ensure that new collaborators have the necessary tools and directions to produce the intended physics selections.  Additionally, these trainings also provide clear guidelines and directions to PWG liaisons to facilitie a streamlined campaign workflow.  Campaigns are broken down into three stages: (1) Development, (2) Validation, and (3) Data Production.

\subsubsection{Development Stage}

During the development stage of the campaign, analysts and maintainers primarily work within the GitLab project. Analysts develop their modules within designated packages, submitting MRs to their respective branches. The corresponding PWG liaison oversees these MRs, ensuring they are correctly named, linked to relevant GitLab Issues and Milestones, and assigned appropriate labels and reviewers. This structured approach helps maintain clear tracking of ongoing developments and facilitates efficient project management.

The liaison is also responsible for testing each MR to ensure the functionality of Stripping lines, verifying that performance metrics remain within predefined thresholds.  The CI system ensures that new code contributions remain compatible with the existing framework by running functional tests on physics selections. It monitors algorithm performance, including rates and timing, to maintain consistency and efficiency across different working groups. For multiple development branches, the CI performs automated code linting and executes standardized tests across all relevant configurations, helping streamline contributions and maintain code quality.

To enable CI tests to access necessary data, appropriate permissions must be granted through the designated analysis production system. Once set up, the CI pipeline compiles a compatible version of the analysis framework using the selected platform and container, integrating key software packages. Branch naming schemes are used to steer different subjobs that are assigned to specific working groups, ensuring targeted execution of tests and validation processes while not overloading CI resources. CI tests must pass successfully before an MR is merged. Once merged, the liaison generates line dictionaries for analysts to review, and all PWG branches are consolidated into the production branch on a set schedule. This systematic process ensures that validation begins on time and that the campaign progresses smoothly.

\subsubsection{Validation Stage}

Following the development stage, a time-period is dedicated to testing and validation to confirm that selections and physics behave as expected. In the validation stage, analysts have access to larger datasets to check for bugs or unintended physics selections.  The workflow for desired changes during this period follows exactly that of the development stage.  Development of brand new selection lines during this stage are prohibited in efforts to stay within the scheduled operational timeline.

\subsubsection{Data Production}

The full Stripping production requests are facilitated by the distributed computing team. In an effort to harmonize efforts with ``live" data processing procedures, and the legacy data processing steps, the Stripping requests use the same workflow as data processing requests of the Run 3 offline stage.

\begin{figure}[h]
\centering
\includegraphics[width=.34\linewidth]{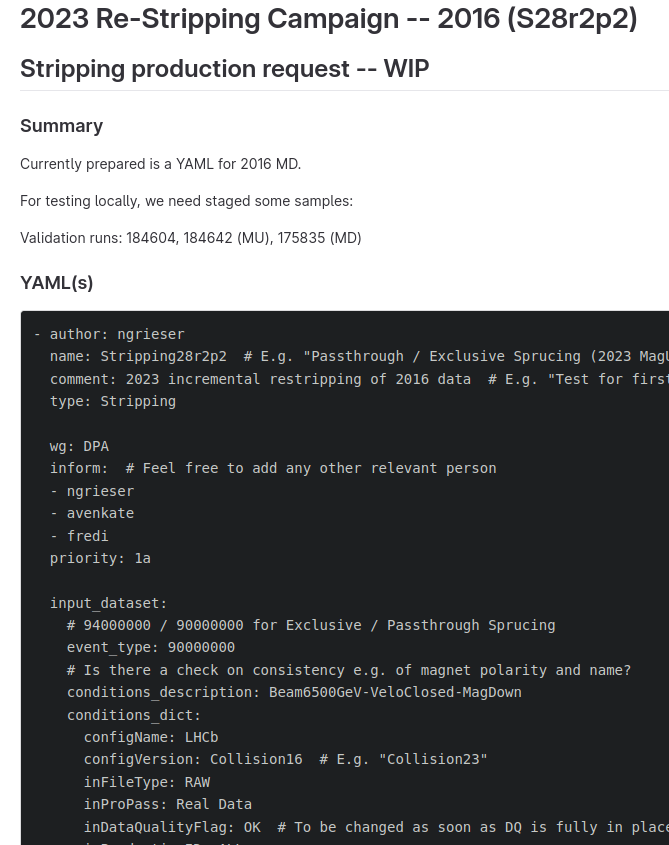}
\includegraphics[width=.63\linewidth]{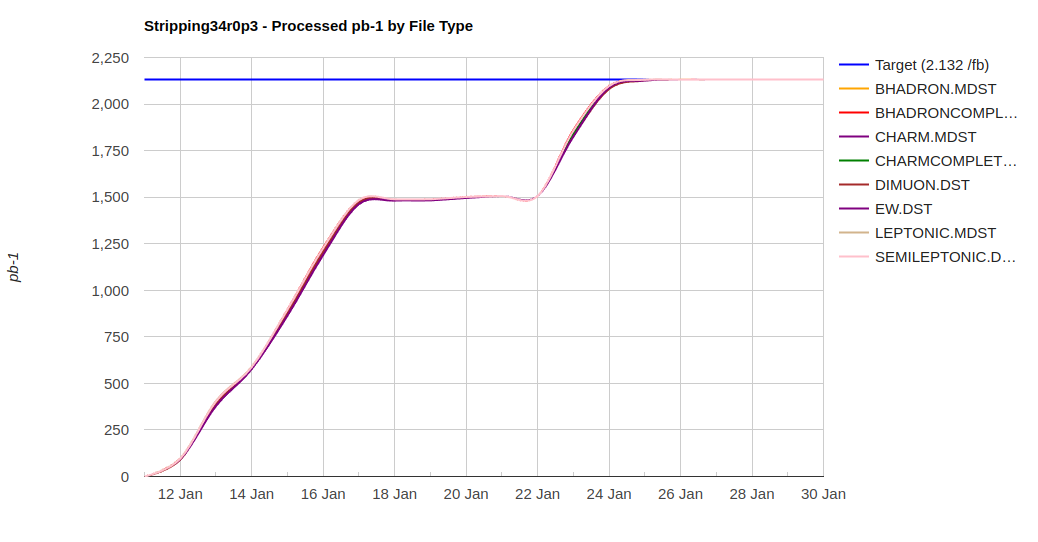}
\caption{Communication between distributed computing teams and the Stripping Coordination is critical to streamline the data processing stage.  A YAML-format file (left) is used for steering the details of the request in conjunction with GitLab Issues.  Monitoring of the production (right) is then provided by the distributed computing team.}
\label{fig:DistComp}       
\end{figure}

GitLab Issues are used to bookkeep data production requests.  YAML files are used to manage Dirac\footnote{Dirac is the LHCb Distributed Computing workload and data management system.} \cite{Dirac} requests in Stripping, defining issues for each production step to request data staging when necessary. These files contain essential processing information, such as years, options, detector conditions, data packages, and applications, allowing for easy reuse and organization.  During the production stage, the distributed computing team provides live updates of production progress.  The Stripping Coordination team then provides regular updates for management teams and analysts of the on-going process.  An example of one of the request YAML files and a production progress plot are provided in Figure~\ref{fig:DistComp}.

\section{Continuously Learning From Campaigns}

\begin{figure}[hbtp]
\centering
\includegraphics[width=.99\linewidth]{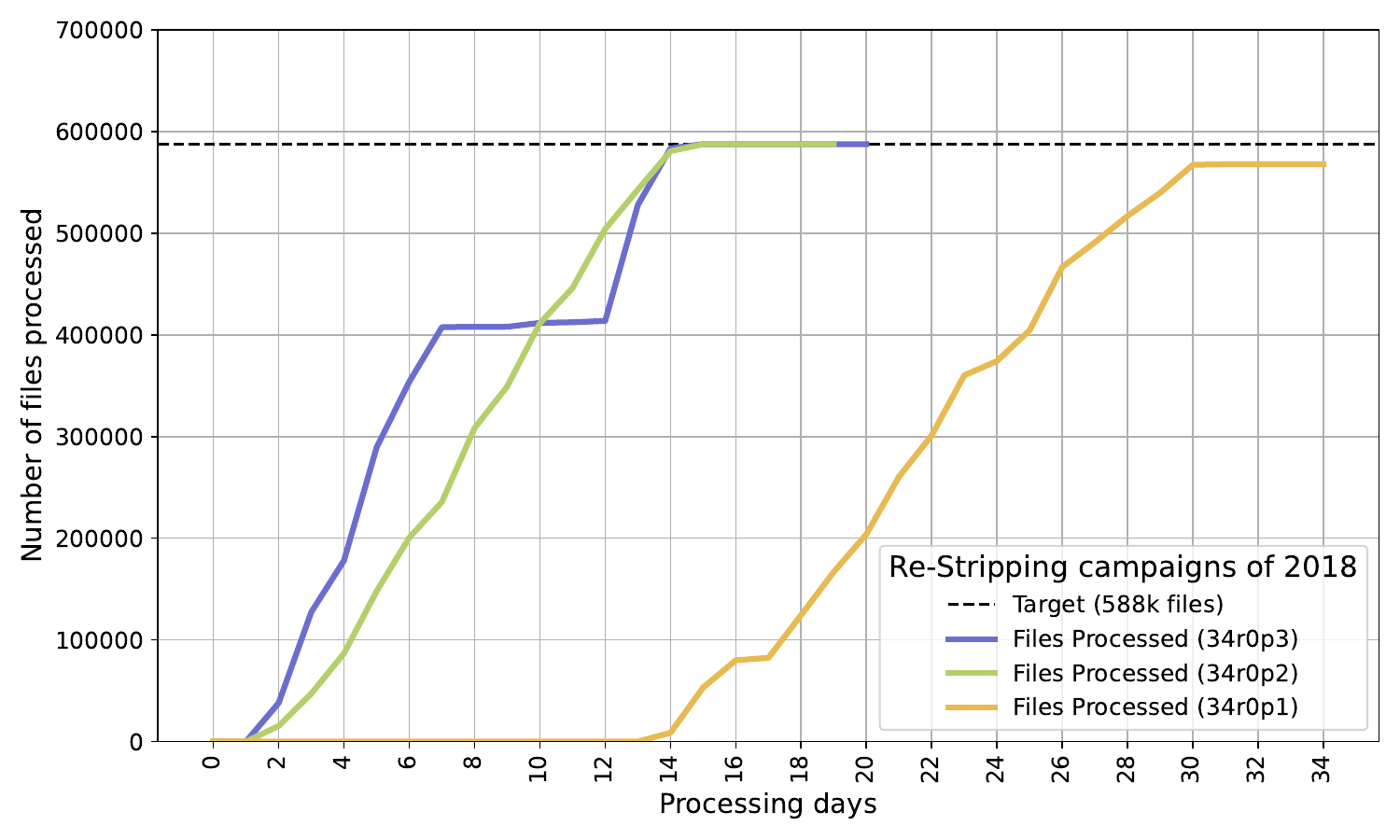}
\caption{A comparison of the processing times of the LHCb 2018 dataset for different Stripping campaigns (34r0p1 being the earliest, and 34r0p3 being the most recent).  Workflow improvements and streamlining efforts are shown to have significant impact in operational time requirements.}
\label{fig:Progression}       
\end{figure}

Legacy data software stacks are rarely being updated with modern developments in a HEP ecosystem that is commonly focused on the most recent datasets.  However, when re-processing older data is possible, applying modern workflows and collaborative ideas can help reach the operational constraints of experiments.

Following large Stripping campaigns, a post-mortem is critical to continuing progress on the procedure.  As legacy data productions begin to be sporadic in schedule, such productions often face high turnover of participants. While challenging, these changes have the opportunity to introduce fresh perspectives and ideas. To manage this effectively, it’s essential to ensure that training and procedures are continuously assessed and improved over time.  For example, following the most recent Stripping campaign a survey of PWG liaisons was distributed.  The evaluations showed that targeted training significantly boosts participants’ comfort and qualifications in their roles. Additionally, the impact of training on role confidence and its contribution to broader professional development were considered. The goal is to ensure that new team members are well-prepared and able to grow in their physics career following responsibilities carried out in legacy campaigns.  By  considering the training value of the responsibilities, future coordination teams will be able to more easily attract participants to legacy data production campaigns in the future.

Moving beyond the human aspect, it is also critical to monitor the impact of workflow changes on the operational performance of these large data re-processing campaigns.  Visualization of the progress of data production time is provided in Figure~\ref{fig:Progression}.  Evidence supports that applying modern workflows and ensuring good communication across activities leads to an increase in efficiency of data production campaigns.

\section{Conclusions}

The LHCb collaboration continues to maintain a thriving legacy physics program, primarily utilizing the Run 1 and Run 2 datasets even as the focus shifts toward live Run 3 data. To ensure analysts can still benefit from these valuable legacy datasets, sustainable and efficient systems are maintained through ongoing software development and large-scale data reprocessing campaigns. The LHCb Stripping project plays a critical role, allowing researchers to efficiently select physics candidates using a flexible Python-based framework. Successful collaboration with computing and operations teams, along with the use of modern tools like GitLab Milestones and continuous integration, helps manage the complexity of large campaigns and maintain software reliability. As workflows evolve, the collaboration continues to foster growth and learning opportunities for both experienced and junior researchers alike.

%
 \bibliography{main.bib} 
%
%
%
%

\end{document}